\documentclass{emulateapj}

\begin{document}

\shortauthors{Luhman \& Esplin}
\shorttitle{Parallax of the Coldest Known Brown Dwarf}

\title{A New Parallax Measurement for the Coldest Known Brown Dwarf\altaffilmark{1}}

\author{K. L. Luhman\altaffilmark{2,3} and T. L. Esplin\altaffilmark{2}}

\altaffiltext{1}
{Based on observations made with the {\it Spitzer Space Telescope}, which is
operated by the Jet Propulsion Laboratory, California
Institute of Technology under a contract with NASA.}

\altaffiltext{2}{Department of Astronomy and Astrophysics,
The Pennsylvania State University, University Park, PA 16802, USA;
kluhman@astro.psu.edu}
\altaffiltext{3}{Center for Exoplanets and Habitable Worlds, The 
Pennsylvania State University, University Park, PA 16802, USA}

\begin{abstract}

WISE J085510.83$-$071442.5 was recently discovered as the coldest known
brown dwarf based on four epochs of images from the {\it Wide-field Infrared
Survey Explorer} and the {\it Spitzer Space Telescope}.
We have improved the accuracy of its parallax measurement by obtaining two
additional epochs of {\it Spitzer} astrometry.
We derive a parallactic distance of 2.31$\pm$0.08~pc, which continues
to support its rank as the fourth closest known system to the Sun when
compared to WISE J104915.57$-$531906.1~AB (2.02$\pm$0.02~pc) and Wolf 359
(2.386$\pm$0.012~pc).
The new constraint on the absolute magnitude at 4.5~\micron\ indicates
an effective temperature of 235--260~K based on four sets of theoretical models.
We also show the updated positions of WISE J085510.83$-$071442.5 in two
color-magnitude diagrams. 
Whereas Faherty and coworkers cited its location in $M_{W2}$ versus $J-W2$
as evidence of water clouds, we find that those data can be explained 
instead by cloudless models that employ non-equilibrium chemistry.

\end{abstract}


\section{Introduction}

Distance is a key parameter in characterizing the physical properties of
brown dwarfs and testing models of their atmospheres and interiors.
A distance estimate enables the measurement of absolute magnitudes
in various photometric bands. The spectral energy distribution constructed
from those magnitudes can be compared to theoretical predictions
in order to derive stellar parameters like mass and effective temperature
and to discriminate among competing models.
Distances of nearby L and T dwarfs have been measured via trigonometric
parallaxes through imaging at red optical and near-infrared (IR) wavelengths
\citep{dah02,tin03,vrb04,mar10,and11,fah12,dup12,man13,sma13,wan14,zap14}.
Because near- to mid-IR colors become rapidly redder toward the end of the
T spectral sequence \citep{kir11}, parallaxes of late-T and Y dwarfs
\citep{tin12,bei13,bei14,dup13,mar13} have required the most sensitive
near-IR cameras that are available, namely those on 8--10~m ground-based
telescopes and the {\it Hubble Space Telescope}, or the mid-IR cameras on the
{\it Wide-field Infrared Survey Explorer} \citep[{\it WISE},][]{wri10} and
the {\it Spitzer Space Telescope} \citep{wer04}.

WISE J085510.83$-$071442.5 (hereafter WISE 0855$-$0714)
is a recently discovered brown dwarf for which a
parallax measurement has had important implications.
It was identified as a high proper motion object by \citet{luh14a}
based on two epochs of images from {\it WISE}\footnote{\citet{kir14}
also reported it as a high proper motion object in a later study.}.
By obtaining two additional epochs of astrometry with {\it Spitzer},
\citet{luh14b} confirmed its large proper motion and measured its parallax.
The parallactic distance of $2.20^{+0.24}_{-0.20}$~pc was roughly midway
between the distances of the third and fourth closest neighbors that were
previously known,
WISE J104915.57$-$531906.1~AB \citep[2.02$\pm$0.02~pc,][]{luh13,bof14} and
Wolf 359 \citep[2.386$\pm$0.012~pc,][]{wol19,van95}.
Based on its absolute magnitude at 4.5~\micron\ and its $[3.6]-[4.5]$ color,
\citet{luh14b} also found that WISE 0855$-$0714 was the coldest known
brown dwarf ($T_{\rm eff}\sim225$--260~K).
\citet{wri14} measured a fifth epoch of astrometry with the reactivated
{\it WISE} satellite \citep[{\it NEOWISE},][]{mai14} and derived new
estimates of the proper motion and parallax.
In this paper, we present two new epochs of {\it Spitzer} astrometry
for WISE 0855$-$0714, which are used to improve the accuracy of
its parallax measurement.

\section{{\it Spitzer} Astrometry}
\label{sec:irac}

\begin{figure}
\epsscale{1.1}
\plotone{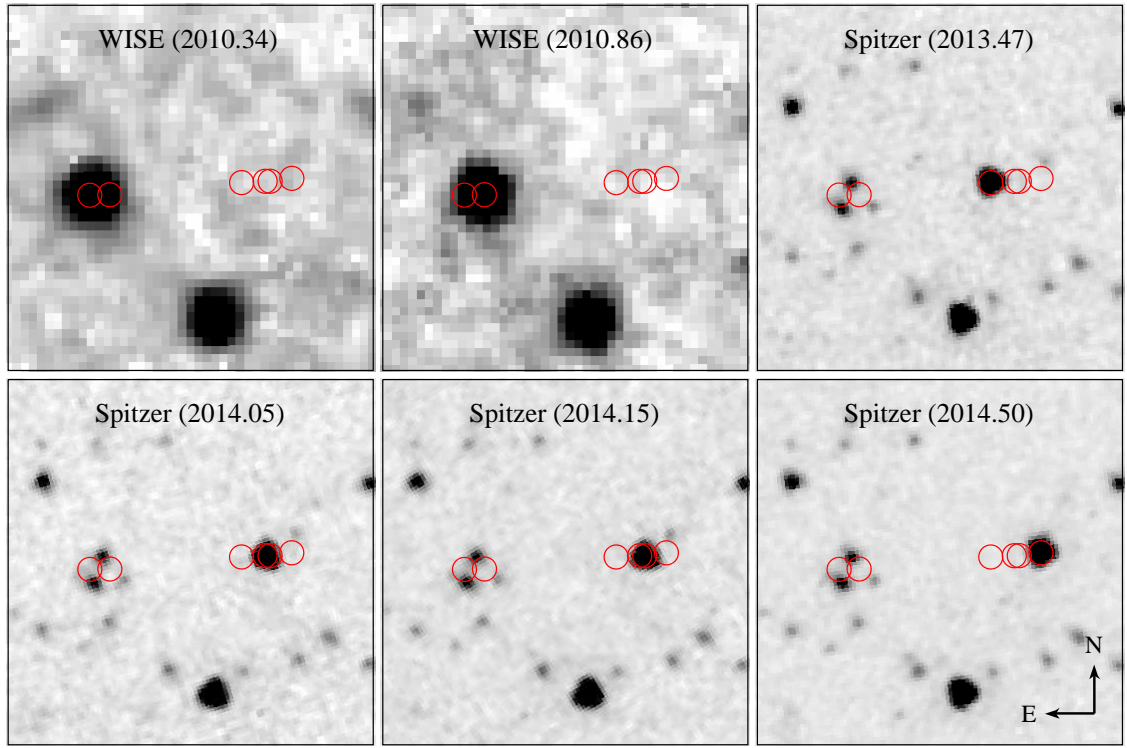}
\caption{
Images of WISE~0855$-$0714 from {\it WISE} ($W2$) and {\it Spitzer} ([4.5]).
The first four images were presented by \citet{luh14b} and the
latest two images were obtained in this work.
The positions of WISE~0855$-$0714 are marked by the circles.
The size of each image is $1\arcmin\times1\arcmin$.
}
\label{fig:image}
\end{figure}

\citet{luh14b} obtained images of WISE 0855$-$0714 with {\it Spitzer}'s
Infrared Array Camera \citep[IRAC;][]{faz04} on 2013 June 21 and
2014 January 20. To further refine its parallax measurement, we observed
WISE 0855$-$0714 with IRAC on two additional dates,
February 24 and July 1 in 2014.
These observations were performed through Astronomical Observation Requests
49096192 and 51040000 within programs 90095 and 10168, respectively.
IRAC currently operates with filters centered at 3.6 and 4.5~\micron, which are
denoted as [3.6] and [4.5]. Only the [4.5] filter was selected for
our imaging because it offers much better sensitivity to cold brown dwarfs.
The camera has a plate scale of $1\farcs2$~pixel$^{-1}$ and 
a field of view of $5\farcm2\times5\farcm2$. It produces images with
FWHM$=1\farcs7$. The exposure time for the individual frames was 26.8~s.
Five and nine dithered frames were collected during the observations
in February and July, respectively.
The two reduced images are shown in Figure~\ref{fig:image} 
with the previous epochs of {\it WISE} and {\it Spitzer} data from
\citet{luh14b} for a $1\arcmin\times1\arcmin$ area surrounding WISE~0855$-$0714.

We have measured astrometry for WISE~0855$-$0714 in each of the four epochs
of {\it Spitzer} data in the following manner, which produces more accurate
results than the methods applied to the first two epochs by \citet{luh14b}.
Pixel coordinates were measured for all point sources in the corrected
basic calibrated data (CBCD) versions of the individual [4.5] exposures
using the Astronomical Point-source
Extractor (APEX) Single Frame pipeline within the Mosaicking and Point-source
Extraction software package \citep{mak05}.
We applied a new distortion correction to those pixel coordinates that
is more accurate than the one that is available from the {\it Spitzer} 
pipeline \citep[T. Esplin, in preparation; see also][]{dup13}.
The corrected coordinates for stars
with S/N$>20$ and detections in more than two frames were used to compute
relative offsets in x, y, and rotation among the frames for a given epoch.
We used the APEX multiframe pipeline to combine the registered CBCD images
for each epoch and measure the positions of all detected sources.
We identified all objects from the Point Source Catalog of
the Two Micron All-Sky Survey \citep[2MASS,][]{skr06} that were within
$2\arcmin$ of WISE~0855$-$0714, were not blended with other stars in the images
from {\it Spitzer}, 2MASS, and {\it WISE}, and have proper motions of
$\lesssim0\farcs02$~yr$^{-1}$, which resulted in a sample of 15 stars.
We measured offsets in right ascension, declination,
and rotation for that sample between 2MASS and the first {\it Spitzer} epoch
and applied them to the source catalog from the latter to
align it to the 2MASS astrometric system. The catalogs from the
other {\it Spitzer} epochs were then aligned to the first epoch
using stars that were within $2\arcmin$ of WISE~0855$-$0714 and have $[4.5]<17$.
To characterize the errors in the astrometry for WISE 0855$-$0714, 
we first computed the differences in right ascension and declination
between adjacent IRAC epochs for stars in a magnitude range encompassing
WISE 0855$-$0714 ($[4.5]=13$--16).
We then estimated the 1~$\sigma$ errors based on two statistics,
the median absolute deviation of those differences and the deviations
from zero that contained 68\% of the distribution.
The astrometry for WISE~0855$-$0714 from each of the four {\it Spitzer} epochs
is presented in Table~\ref{tab:astro}.

\section{{\it WISE} Astrometry}
\label{sec:wise}

In addition to the astrometry from {\it Spitzer}, we also make use of the
detections of WISE~0855$-$0714 from {\it WISE} and {\it NEOWISE} when
measuring its parallax. 
By comparing the {\it WISE} images to the first two epochs from {\it Spitzer},
\citet{luh14b} found that WISE~0855$-$0714 is
blended with a group of background objects (primarily two sources with similar
fluxes) in both epochs from {\it WISE} (see Figure~\ref{fig:image}).
For each of the {\it WISE} epochs, \citet{luh14b} measured astrometry
for the blend of WISE~0855$-$0714 and the contaminants from the coaddition of
the single-exposure images at 4.6~\micron\ (denoted as $W2$), which
is the {\it WISE} band in which WISE~0855$-$0714 dominates.
Because WISE~0855$-$0714 had moved away from the background sources by the
time of the {\it Spitzer} observations, \citet{luh14b} was able to use the 
{\it Spitzer} images to estimate the true positions of WISE~0855$-$0714 in the
{\it WISE} epochs in the following manner. For a grid of locations in a
{\it Spitzer} image surrounding the {\it WISE} coordinates of WISE~0855$-$0714, 
he added an artificial star with the [4.5] flux of WISE~0855$-$0714, smoothed
the image to the resolution
of {\it WISE}, and measured astrometry for the blend of the artificial star
and the background objects. For the simulated image in which the latter
coordinates matched the astrometry measured for the blend in {\it WISE}, the
artificial star's inserted location was adopted as the true position of
WISE~0855$-$0714.
\citet{wri14} measured new astrometry for the blend of WISE~0855$-$0714
and its contaminants by applying software developed for the AllWISE
Source Catalog to the $W2$ images from each of the two {\it WISE} epochs.
They also measured astrometry for the brown dwarf from $W2$ images
that were obtained in May of 2014 by {\it NEOWISE}. 
To account for the blending in the {\it WISE} images, \citet{wri14}
included a parameter that related the true position to the observed position
in their least-squares fitting of the proper motion and parallax.

For our analysis, we adopt the astrometry measured by \citet{wri14} for
the blends of WISE~0855$-$0714 and the background sources in the two 
{\it WISE} epochs and for the brown dwarf alone in the {\it NEOWISE} epoch.
To place those data on the same astrometric system as the {\it Spitzer}
astrometry, we calculated the average differences in right ascension and
declinations between 2MASS and AllWISE for the 15 2MASS reference stars
from Section~\ref{sec:irac}. The resulting offset of (0.1, $0\arcsec$)
was added to ($\alpha$, $\delta$) for each of the three epochs of astrometry
from \citet{wri14}. We are assuming that those data from \citet{wri14}
are on the same astrometric system as the AllWISE Source Catalog.
To correct the astrometry from {\it WISE} for the blended background sources,
we applied the procedure from \citet{luh14b} that was summarized earlier
in this section. 
For the errors in the corrected positions, we adopted the ranges in right
ascensions and declinations of the inserted artificial stars that reproduced
the errors in the observed, blended astrometry from \citet{wri14}.
We prefer the method of correcting the astrometry from \citet{luh14b}
over the one from \citet{wri14} because the former makes use of the
accurate astrometry and photometry of the contaminants that are available from
{\it Spitzer}, which should comprise all of the information necessary for a
reliable simulation of the blending.
In addition, the fitting procedure from \citet{wri14} could produce erroneous
results if astrometric perturbations from a companion are present.
Table~\ref{tab:astro} contains the final astrometry that we adopt for
WISE~0855$-$0714 from {\it WISE} and {\it NEOWISE}.

\section{Parallax and Proper Motion}
\label{sec:astro}

\subsection{Previous Measurements}

Two previous studies, \citet{luh14b} and \citet{wri14}, have reported
measurements of the proper motion ($\mu$) and parallax ($\pi$) of
WISE~0855$-$0714. Based on the two epochs from {\it WISE} and the first
two epochs from {\it Spitzer}, \citet{luh14b} arrived at 
$(\mu_{\alpha},\mu_{\delta})=(-8.06\pm0.09,0.70\pm0.07\arcsec$~yr$^{-1}$)
and $\pi=0.454\pm0.045\arcsec$. 
\citet{wri14} added a fifth epoch of astrometry from {\it NEOWISE}, measured
new astrometry from the {\it WISE} epochs, and addressed the blending
in the latter with an alternative method from that in \citet{luh14b},
as discussed in Section~\ref{sec:wise}. Combining those data with
the {\it Spitzer} astrometry from \citet{luh14b}, they derived
$(\mu_{\alpha},\mu_{\delta})=(-8.051\pm0.047,0.657\pm0.050\arcsec$~yr$^{-1}$)
and $\pi=0.448\pm0.033\arcsec$.
Their errors in the proper motion and parallax were smaller than those
from \citet{luh14b} because of the additional epoch from {\it NEOWISE} and
the smaller errors for their {\it WISE} astrometry.

\citet{wri14} concluded that their analysis confirmed the results from
\citet{luh14b}. However, \citet{wri14} did not present new astrometry that
was capable of accurately detecting a large parallax independently from the
{\it Spitzer} images.
Instead, rather than confirming the large parallax, their one additional
epoch refined the proper motion, which in turn allowed the {\it Spitzer} data 
to slightly better constrain the parallax.
\citet{wri14} also demonstrated that an alternative correction for the 
contamination of WISE~0855$-$0714 in the {\it WISE} images
produces a similar proper motion and parallax as the correction method
from \citet{luh14b}.

\subsection{New Measurements}

\begin{figure}
\epsscale{1.2}
\plotone{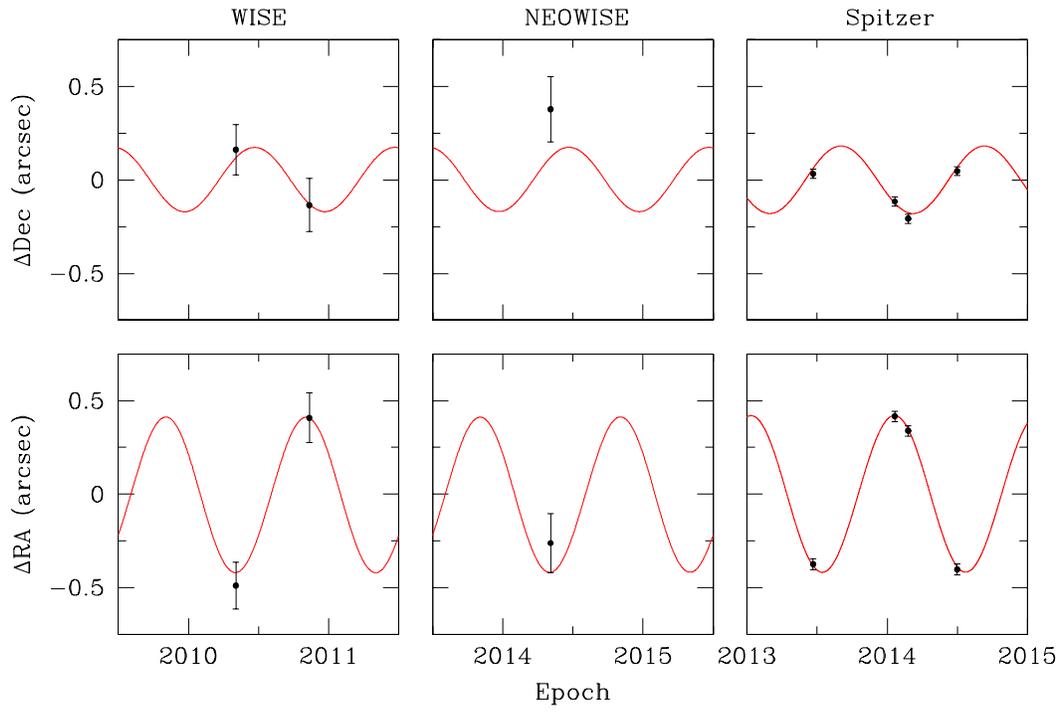}
\caption{
Relative astrometry of WISE~0855$-$0714 (Table~\ref{tab:astro}) compared
to the best-fit model of parallactic motion (Table~\ref{tab:data}, red curve).
The proper motion produced by the fitting has been subtracted.
}
\label{fig:pm}
\end{figure}

We applied least-squares fitting of proper and parallactic motion to the
seven epochs of astrometry for WISE~0855$-$0714 in Table~\ref{tab:astro}
with the IDL program MPFIT.
The reduced $\chi^2$ was fairly close to unity (0.5), indicating a good fit.
We checked the errors from that procedure by creating 1000 sets of
astrometry that consisted of the sum of the measured astrometry and Gaussian
noise, and fitting parallactic and proper motion to each set.
The resulting standard deviations of $\mu_{\alpha}$ and $\mu_{\delta}$
were similar to the errors from MPFIT.
However, the standard deviation of the parallax was larger
than the MPFIT error ($0\farcs015$ vs.\ $0\farcs013$); we have adopted the
former for the parallax error. Our derived proper motion and
parallax are presented in Table~\ref{tab:data}. They are consistent with
the previous estimates from \citet{luh14b} and \citet{wri14}.
We show the relative coordinates among the seven epochs in Figure~\ref{fig:pm}
after subtraction of the best-fit proper motion.

\section{Discussion}

The original parallactic distance of $2.20^{+0.24}_{-0.20}$~pc
for WISE~0855$-$0714 \citep{luh14b} was $\sim1$~$\sigma$ from
the distances of WISE J104915.57$-$531906.1~AB
\citep[2.02$\pm$0.02~pc,][]{bof14} and
Wolf 359 \citep[2.386$\pm$0.012~pc,][]{van95}, which were the third
and fourth closest systems to the Sun that were known prior to the discovery of
WISE~0855$-$0714.
Our measurement of 2.31$\pm$0.08~pc demonstrates more definitively
that WISE~0855$-$0714 is likely the fourth closest known system.
Among known Y dwarfs, WISE~0855$-$0714 is now roughly tied with WD~0806-661~B
\citep{sub09} for the smallest percentage error in its parallax ($\sim3$--4\%).

\citet{luh14b} estimated the effective temperature of WISE~0855$-$0714
by comparing $M_{4.5}$ and a limit on $J-[4.5]$ to the values predicted
by atmospheric and evolutionary models of brown dwarfs. 
The selected models were defined primarily by the following features:
water clouds and chemical equilibrium \citep{bur03},
cloudless and chemical equilibrium \citep{sau12},
cloudless and non-equilibrium chemistry \citep{sau12}, and
50\% coverage of water, chloride, and sulfide clouds and chemical equilibrium
\citep{mor12,mor14}.
The latter three sets of atmospheric models utilize the evolutionary
calculations of \citet{sau08}.
The cloudless models with equilibrium and non-equilibrium chemistry
from \citet{sau12} are the same as those used by \citet{luh12} and \citet{luh14b}.
Cloudy models with non-equilibrium chemistry are also available \citep{mor14},
but they were not considered since they predict nearly identical photometry
in $J$ and [4.5] as the cloudy equilibrium models for the coldest Y dwarfs.
When we repeat the temperature estimates from \citet{luh14b} using our
parallax measurement and the $J$-band photometry from \citet{fah14}, we find
that the new constraints on $J-[4.5]$ and $M_{4.5}$ imply temperatures
of 225--280~K and 235--260~K, respectively, based on the four sets of
models, and 225--240~K and 250--260~K for the models of \citet{mor14}.

\begin{figure}
\epsscale{1}
\plotone{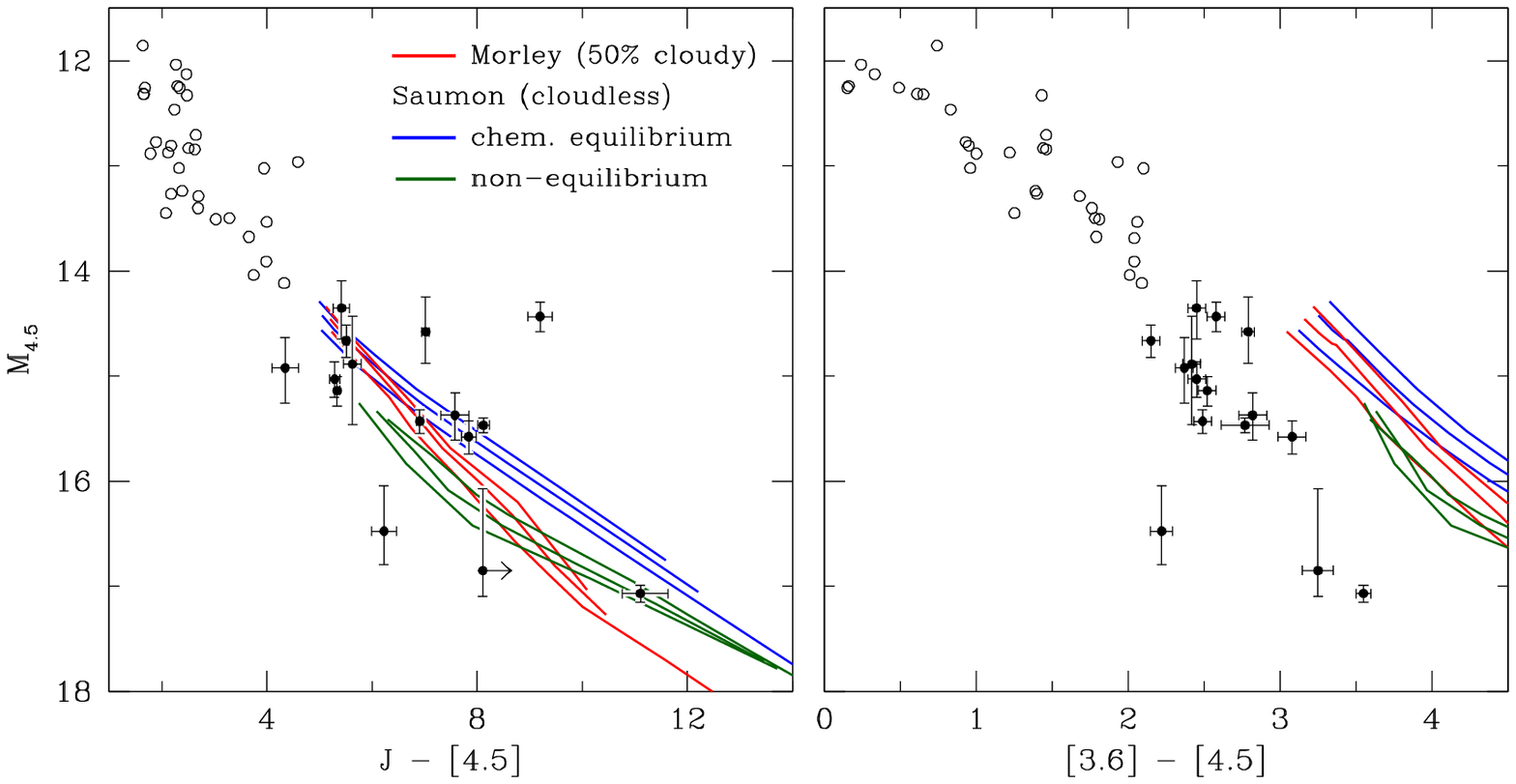}
\caption{
Color-magnitude diagrams for WISE~0855$-$0714
\citep[faintest point,][this work]{luh14b,fah14} and 
samples of T dwarfs \citep[open circles,][references therein]{dup12}
and Y dwarfs \citep[filled circles with error
bars,][]{cus11,cus14,tin12,luh12,luh14wd,bei13,bei14,leg13,mar13,kir13,dup13}.
These data are compared to the magnitudes and colors predicted
by theoretical models for ages of 1, 3, and 10~Gyr
\citep[solid lines,][]{sau12,mor12,mor14}.
}
\label{fig:cmd}
\end{figure}

Previous studies have compared WISE~0855$-$0714 to other Y dwarfs and
to theoretical models via color-magnitude diagrams \citep{luh14b,luh14wd,fah14}.
In Figure~\ref{fig:cmd}, we show the positions of WISE~0855$-$0714
in diagrams of $M_{4.5}$ versus $J-[4.5]$ and $M_{4.5}$ versus $[3.6]-[4.5]$
based on our parallax measurement and photometry from \citet{luh14b} and
\citet{fah14} (see Table~\ref{tab:data}).
We have selected [4.5] as the magnitude since it offers the smallest
photometric errors among the filters in which WISE~0855$-$0714
and other Y dwarfs have been observed.
For comparison, we have included in those diagrams data for known T and
Y dwarfs with measured parallaxes and photometry in $J$, [3.6], and [4.5]
\citep{cus11,cus14,dup12,tin12,luh12,luh14wd,bei13,bei14,leg13,mar13,kir13,dup13}\footnote{The uncertainty in the estimate of $J$ for WD~0806-661~B
from \citet{luh14wd} is not well-determined, but we have adopted a value of
0.1~mag for Figure~\ref{fig:cmd}.}.
WISE~0855$-$0714 now has the best constraint
on its position in $M_{4.5}$ versus $[3.6]-[4.5]$ among known Y dwarfs.

In addition to the data for T and Y dwarfs, we also plot in
Figure~\ref{fig:cmd} the magnitudes and colors predicted by three of the
four sets of models described earlier. We have omitted the models of
\citet{bur03} since they differ the most from the data for WISE~0855$-$0714
(see \citealt{luh14wd}).
Among the remaining models, those using equilibrium and non-equilibrium
chemistry are shown for temperatures of $<450$~K and $<350$~K, respectively.
Except for WD~0806-661~B, the ages of the known Y dwarfs are unknown, so
we have shown the model predictions for ages of 1, 3, and 10~Gyr, which
span the ages of most stars in the solar neighborhood. 
As discussed previously \citep{leg10a,bei14,luh14wd},
these theoretical isochrones are significantly redder than the data for
WISE~0855$-$0714 and other T and Y dwarfs in $M_{4.5}$ versus $[3.6]-[4.5]$.
In $M_{4.5}$ versus $J-[4.5]$, WISE~0855$-$0714 is 1.5~$\sigma$ bluer
than the cloudless/chemical equilibrium models, 2.5~$\sigma$ redder than
the cloudy models, and agrees with the cloudless/non-equilibrium models.
For the Y dwarf sequence as a whole, no one set of
models provides a clearly superior match.

In their analysis of a diagram of $M_{W2}$ versus $J-W2$, which is
analogous to  $M_{4.5}$ versus $J-[4.5]$,
\citet{fah14} found that WISE~0855$-$0714 was 2.7~$\sigma$ bluer than
the cloudless/chemical equilibrium models from \citet{sau12} and was within
1~$\sigma$ of the cloudy models from \citet{mor12,mor14}, which they cited
as the first evidence of water ice clouds outside the solar system.
However, WISE~0855$-$0714 is closer to those cloudless models than the
cloudy models in $M_{4.5}$ versus $J-[4.5]$, as shown in Figure~\ref{fig:cmd}.
Furthermore, in both $M_{4.5}$ versus $J-[4.5]$ and $M_{W2}$ versus $J-W2$,
the positions of WISE~0855$-$0714 are reproduced by the cloudless models
from \citet{sau12} that use non-equilibrium chemistry.
Thus, those data do not serve as evidence for or against the presence of
water clouds.
Nevertheless, WISE~0855$-$0714 is expected to contain such clouds given
that they are predicted to form at $T_{\rm eff}<400$~K \citep{bur03,mor14}.

\acknowledgements
We acknowledge support from grant NNX12AI47G from the NASA Astrophysics
Data Analysis Program. We thank Caroline Morley and Didier Saumon for
providing their model calculations.
{\it WISE} is a joint project of the University of California, Los Angeles,
and the Jet Propulsion Laboratory (JPL)/California Institute of
Technology (Caltech), funded by NASA.
2MASS is a joint project of the University of
Massachusetts and the Infrared Processing and Analysis Center (IPAC) at
Caltech, funded by NASA and the NSF.
The Center for Exoplanets and Habitable Worlds is supported by the
Pennsylvania State University, the Eberly College of Science, and the
Pennsylvania Space Grant Consortium.

\clearpage

\begin{deluxetable}{llllll}
\tabletypesize{\scriptsize}
\tablewidth{0pt}
\tablecaption{Astrometry for WISE~J085510.83$-$071442.5\label{tab:astro}}
\tablehead{
\colhead{$\alpha$ (J2000)} & \colhead{$\sigma_{\alpha}$} & \colhead{$\delta$ (J2000)} & \colhead{$\sigma_{\delta}$} & \colhead{MJD} &  \colhead{Source} \\
\colhead{($\arcdeg$)} & \colhead{($\arcsec$)} & \colhead{($\arcdeg$)} & \colhead{($\arcsec$)} & \colhead{} & \colhead{}}
\startdata
133.7952573 & 0.125 & $-$7.2450910 & 0.135 & 55320.4 & {\it WISE} \\
133.7943232 & 0.133 & $-$7.2450719 & 0.142 & 55511.3 & {\it WISE} \\
133.7881873 & 0.028 & $-$7.2445207 & 0.024 & 56464.5 & {\it Spitzer} \\
133.7870881 & 0.028 & $-$7.2444491 & 0.024 & 56677.3 & {\it Spitzer} \\
133.7868488 & 0.028 & $-$7.2444561 & 0.024 & 56712.3 & {\it Spitzer} \\
133.7862461 & 0.158 & $-$7.2442562 & 0.175 & 56782.4 & {\it NEOWISE} \\
133.7858505 & 0.028 & $-$7.2443176 & 0.024 & 56839.7 & {\it Spitzer} \\
\enddata
\tablecomments{The {\it WISE} and {\it NEOWISE} data are from \citet{wri14}
after the adjustments described in Section~\ref{sec:wise}.}
\end{deluxetable}

\begin{deluxetable}{lll}
\tabletypesize{\scriptsize}
\tablewidth{0pt}
\tablecaption{Parallax, Proper Motion, and Photometry for
WISE~J085510.83$-$071442.5\label{tab:data}}
\tablehead{
\colhead{Parameter} & \colhead{Value} & Reference 
}
\startdata
$\pi$ & $0.433\pm$0.015$\arcsec$ & 1 \\
$\mu_{\alpha}$ cos $\delta$ & $-8.10\pm0.02\arcsec$~yr$^{-1}$ & 1 \\
$\mu_{\delta}$ & $0.70\pm0.02\arcsec$~yr$^{-1}$ & 1 \\
$Y$ & $>24.4$\tablenotemark{a} & 2 \\
$J$ & $25.0^{+0.53}_{-0.35}$ & 3 \\
$H$ & $>22.7$\tablenotemark{a} & 4 \\
$K_s$ & $>18.6$\tablenotemark{a} & 5 \\
$W1$ & 17.82$\pm$0.33 & 4 \\
$W2$ & 14.02$\pm$0.05 & 4 \\
$[3.6]$ & 17.44$\pm$0.05 & 6 \\
$[4.5]$ & 13.89$\pm$0.02 & 6 \\
\enddata
\tablenotetext{a}{S/N$<$3.}
\tablerefs{
(1) this work;
(2) \citet{bea14};
(3) \citet{fah14};
(4) \citet{wri14};
(5) VISTA Hemisphere Survey;
(6) \citet{luh14b}.}
\end{deluxetable}

\clearpage

\end{document}